\documentclass[12pt]{amsart}

\usepackage{amssymb,amsfonts,amsthm,multicol,amscd}
\usepackage{amsmath,amsthm,amscd}
\usepackage[mathscr]{eucal}
\allowdisplaybreaks

\usepackage[curve,matrix,arrow]{xy}

\renewcommand{\atop}[2]{\genfrac{}{}{0pt}{}{#1}{#2}}

\newtheorem{Thm}{Theorem}[section]
\newtheorem{Lemma}[Thm]{Lemma}
\newtheorem{Prop}[Thm]{Proposition}
\newtheorem{cor}[Thm]{Corollary}
\newtheorem{rem}[Thm]{Remark}

\theoremstyle{definition}
\newtheorem{Dfn}[Thm]{Definition}

\newcommand{\theA}{\mathcal{A}}
\newcommand{\vect}[1]{{\mathbf{#1}}}

\newcommand{\half}{\frac{1}{2}}

\def\be            {\begin{equation}}
\def\bearl         {\begin{array}{l}}
\def\bearll        {\begin{array}{ll}}
\def\bearlll       {\begin{array}{lll}}

\def\dd            {\partial}

\def\ee            {\end{equation}}
\def\eE            {{\rm e}}
\def\eear          {\end{array}}

\newcommand\Frac[2]{\mbox{\large$\frac{#1}{#2}$}}
\def\futnote#1     {\,\footnote{~#1}\ }

\def\h             {\mathfrak{h}}

\def\ii            {{\rm i}}

\long\def\labl#1#2 {\label{#1#2}\ifnum\draftcontrol=1
                   \mbox{ }\\[-12 mm]\query{#1#2}\\[5 mm] \fi}
\long\def\Labl#1   {\label{#1}\ee \ifnum\draftcontrol=1
                   \mbox{ }\\[-12 mm]\query{#1}\\[5 mm] \fi}

\newcommand\no[1]  {\,\raisebox{.033em}{\large\bf:}#1
                   \raisebox{.033em}{\large\bf:}\,}

\def\la            {\lambda}

\newcommand{\algW}{\mathcal{W}}
\newcommand{\algA}{\mathcal{A}}
\newcommand{\algB}{\mathcal{B}}

\newcommand{\oC}{\mathbb{C}}
\newcommand{\oN}{\mathbb{N}}

\newcommand{\oZ}{\mathbb{Z}}

\newcommand{\repX}{\mathscr{X}}
\newcommand{\repM}{\mathscr{M}}

\newcommand\ket[1]{|#1\rangle}
\newcommand\version[1] {\ifnum\draftcontrol=1 \typeout{}\typeout{#1}\typeout{}
                   \vskip3mm \centerline{\fbox{{\tt DRAFT -- #1 -- }
                   {\small\draftdate}}} \vskip3mm \fi}

\def\tensor{\otimes}
\def\al{\alpha}
\def\be{\beta}
\def\ch{\mbox{\rm ch}}
\def\T{\otimes}

\begin{document}
%\addtolength{\baselineskip}{4pt}
%\addtolength{\parskip}{2pt}

%%% for DRAFT versions, suppress in definitive version
%\draft
%\version\versionno

%%%%%%%%%%%%%%%%%%%%%%%%%%%%%%%%%%%%%%%%%%%%%%%%%%%%%%%%%%%%%%%%%%%%%%%

\title[Fermionic formulas]{%
  \vspace*{-4\baselineskip}
  \mbox{}\hfill\texttt{\small\lowercase{hep-th}/\lowercase{yymmxxx}}
  \\[\baselineskip]Fermionic formulas for $(1,p)$ logarithmic model characters
  in $\Phi_{2,1}$ quasiparticle realisation}
\author{B. Feigin, E. Feigin, I.Tipunin}
\address{BF: Landau institute for Theoretical Physics,
Chernogolovka, 142432, Russia and \newline Independent University of
Moscow, Russia, Moscow, 119002, Bol'shoi Vlas'evski per., 11}
\email{bfeigin@gmail.com}
\address{EF:
Tamm Theory Division, Lebedev Physics Institute, Russia, Moscow,
119991, Leninski pr., 53 and\newline Independent University of
Moscow, Russia, Moscow, 119002, Bol'shoi Vlas'evski per., 11}
\email{evgfeig@gmail.com}
\address{IT:
Tamm Theory Division, Lebedev Physics Institute, Russia, Moscow,
119991, Leninski pr., 53}
\email{tipunin@gmail.com}

\begin{abstract}
  We give expressions for the characters of $(1,p)$ logarithmic
  conformal field models in the Gordon-type form. The formulas are
  obtained in terms of ``quasiparticles'' that are Virasoro
  $\Phi_{2,1}$ primary fields and generalize the symplectic fermions.
\end{abstract}

\maketitle

\section{Introduction}
In recent times, logarithmic conformal field theories are investigated
from different
directions~\cite{[GK2],[GaberdielKausch3],[Kausch-sympl],[Flohr],[FFHST],
  [FGST2],[K-first],[S]}.  There exists a class of models that are
``extensions'' of minimal models~\cite{[DMS]} by some set of vertex
operators~\cite{[FGST3]}.  The most popular are the so called $(1,p)$
models.

The logarithmic $(1,p)$ models have the central charge
\begin{equation}\label{c-charge}
  c=13-6p - \frac{6}{p}.
\end{equation}
The local chiral algebra of the logarithmic $(1,p)$ models is the
triplet W-al\-gebra studied in~\cite{[K-first],[GK2]}.  We let
$\algW(p)$ denote this algebra. The algebra $\algW(p)$ bears an action
of the $s\ell(2)$ algebra that differentiates OPEs.  The adjunctive
triplet refers to the  fact that $\algW(p)$ is generated by the
fields $W^{\pm,0}(z)$, which are transformed as the spin-1
representation of the $s\ell(2)$.  Moreover, $W^+(z)$ and $W^-(z)$ are
highest and lowest weight vectors of the triplet respectively.  The
fields $W^{\pm,0}(z)$ are three solutions of the equation
\begin{equation}
  \partial^3\Phi(z)+\mathrm{const}_1 :T(z)\partial\Phi(z):
   +\mathrm{const}_2 :\partial T(z)\Phi(z):=0
\end{equation}
for the field $\Phi_{3,1}(z)$.  The vertex operator algebra $\algW(p)$
is an extension by~$\Phi_{3,1}(z)$ of the Virasoro algebra with the
central charge~\eqref{c-charge}. We let~$Vac_p$ denote the vacuum
representation of this Virasoro vertex operator algebra.

The algebra $\algW(p)$ has $2p$ irreducible
representations~$\repX^\pm_{s,p}$ ($1\leq s\leq p$).  These
representations admit the action of the $s\ell(2)$ as well.  The
representations labeled with the superscript~$+$ decompose into a
direct sum of odd dimensional irreducible $s\ell(2)$ representations
and labeled with the superscript~$-$ into a direct sum of even
dimensional.  This leads to the factors $2n+1$ and $2n$ in the
characters~\cite{[Flohr],[FFHST]}
\begin{align}
  \chi^+_{s,p}(q)=&\frac{q^{-\frac{1}{24}}}{\prod_{n=1}^{\infty} (1-q^n)}
   \sum_{n\in\oZ}(2n+1)q^{p(n+\frac{p-s}{2p})^2},\label{chi-pl}\\
  \chi^-_{s,p}(q)=&\frac{q^{-\frac{1}{24}}}{\prod_{n=1}^{\infty} (1-q^n)}
   \sum_{n\in\oZ}(2n)q^{p(-n+\frac{s}{2p})^2},\label{chi-min}
\end{align}
where
$\chi^\pm_{s,p}(q)=\mathrm{Tr}_{\repX^\pm_{s,p}}q^{L_0-\frac{c}{24}}$.
These expressions for the characters can be considered ``bosonic''
formulas because they are obtained from a resolution of the
irreducible module constructed from some modules, which can be
considered $\algW(p)$ Verma modules.

We obtain the ``fermionic'' formulas for characters in terms of
``slightly'' bigger algebra $\algA(p)$ that is an extension of
the Virasoro vertex operator algebra~$Vac_p$ with two solutions
of the equation
\begin{equation}
  \partial^2 \Phi(z)+\mathrm{const}:T(z)\Phi(z): =0
\end{equation}
for the field $\Phi_{2,1}(z)$ with conformal
dimension~$\frac{3p-2}{4}$.  The vertex operator algebra $\algA(p)$
bears the action of $s\ell(2)$ and two fields $a^\pm(z)$ are the
highest and the lowest weight vectors of the $s\ell(2)$
spin-$\frac{1}{2}$ irreducible representation.  The fields $a^\pm(z)$
are two highest weight vectors of the $\algW(p)$-module
$\repX^-_{1,p}$.
%The field $a^\pm(z)$ are two
%components of the Virasoro field $\Phi_{2,1}$ .
We note that $\algA(p)$ is nonlocal vertex operator algebra, which
means that there are exist conformal blocks with $\algA(p)$ fields
that have nontrivial monodromy.  In the $p=2$ case, $a^\pm(z)$
coincide with derivative of the symplectic fermions~\cite{[Kausch]}.
In that case ``nonlocality'' leads to two sectors in one of which
symplectic fermions act with integer and in other with half-integer
modes.  We have a sequence of extensions of vertex operator algebras
\begin{equation}
  Vac_p\hookrightarrow\algW(p)\hookrightarrow\algA(p).
\end{equation}
%where $\mathrm Vir$ is the Virasoro algebra of the minimal model, the
%$\algW(p)$ is the maximal local algebra of the logarithmic model and
%$\algA(p)$ is the nonlocal extension of $\algW(p)$ with the
%field~$\Phi_{2,1}$.
 The algebra $\algA(p)$ has $p$ irreducible
representations $\repX_{s,p}$ ($1\leq s\leq p$). Each irreducible
$\algA(p)$ module as a $\algW(p)$ module decomposes
as~$\repX_{s,p}=\repX^+_{s,p}\oplus\repX^-_{s,p}$.  We set
\begin{equation}\label{A-irr-char}
  \chi_{s,p}(q)=\mathrm{Tr}_{\repX^+_{s,p}\oplus\repX^-_{s,p}}q^{L_0-\frac{c}{24}}
 =\chi^+_{s,p}(q)+\chi^-_{s,p}(q).
\end{equation}

The main result of the paper is formulated as follows.
\begin{Thm}\label{thm:main}
  The characters~\eqref{A-irr-char} can be written in the form
\begin{equation}\label{eq:ferm-char}
  \chi_{s,p}(q)=
   q^{\frac{s^2-1}{4p}+\frac{1-s}{2}-\frac{c}{24}}\,\sum_{n_+,n_-,n_1,\dots,n_{p-1}\ge 0}
  \frac{q^{\half\vect{n}\theA\cdot\vect{n} + \vect{v}_s\cdot\vect{n}}}
       {(q)_{n_+}(q)_{n_-}(q)_{n_1}\dots (q)_{n_{p-1}}}\,,
\end{equation}
where  $\vect{n}=(n_+,n_-,n_1,\dots,n_{p-1})$, $(q)_k=\prod_{i=1}^k (1-q^i)$,
$\theA$ is  the Gordon matrix
\begin{equation}\label{eq:the-matrix}
 \theA=%\tiny
\begin{pmatrix}
     \frac{p}{2} &\frac{p}{2} &  1  &  2  &  3  &\dots&  p-1\\
     \frac{p}{2} &\frac{p}{2} &  1  &  2  &  3  &\dots&  p-1\\
       1  &  1  &  2  &  2  &  2  &\dots&   2 \\
       2  &  2  &  2  &  4  &  4  &\dots&   4 \\
       3  &  3  &  2  &  4  &  6  &\dots&   6 \\
     \dots&\dots&\dots&\dots&\dots&\dots&\dots\\
      p-1 & p-1 &  2  &  4  &  6  &\dots& 2(p-1)
  \end{pmatrix}\,,
\end{equation}
and
\begin{equation}
  \vect{v}_s=(\frac{p-s}{2},\frac{p-s}{2},\underbrace{0,\dots0}_{s-1},
 \underbrace{1,2,\dots,p-s}_{p-s}).
\end{equation}
\end{Thm}

Similar but different fermionic formulas were recently obtained
in~\cite{[FlohrF]}.  We emphasis that the fermionic formulas for
characters depend on the chosen set of ``particles'' in terms of
which the formulas are written. The number of particles is equal to
the order of the matrix~\eqref{eq:the-matrix}, i.e.{} $p+1$ in our
case. First two rows and columns correspond to $a^+(z)$ and $a^-(z)$
and other particles appear in singular terms of the
OPE~$a^+(z)a^-(w)$. We note that the matrix obtained by dropping first
two rows and first two columns from~\eqref{eq:the-matrix} coincides
with the standard Gordon matrix~$2\mathrm{min}(i,j)$. Our
considerations in this paper have many overlaps with the construction
for fermionic formulas of minimal models given in~\cite{[FJKLM]} in
terms of the Virasoro primary field~$\Phi_{2,1}$.  Such a construction
is natural in the corner transfer matrix approach to the RSOS models
and their connection with the Virasoro minimal
models~\cite{[Foda],Foda:1998tx}. In this approach one should consider
nonlocal vertex operator algebras, which are extensions of the
Virasoro algebra by a set of primary fields.  However these nonlocal
vertex operator algebras have treatable theory of representation.

We now briefly describe the way we prove Theorem $\ref{thm:main}$.
We construct a degeneration of the chiral algebra $\algA(p)$ to some
algebra $\bar\algA(p)$ with generators called ``particles''
that satisfy a set of quadratic defining relations. The structure of
these quadratic relations is given by the Gordon matrix
$\ref{eq:the-matrix}$.  Each
irreducible representation of $\algA(p)$ has a $\bar\algA(p)$
representation counterpart that is the cyclic $\bar\algA(p)$
representation with the same character.  The characters of
$\bar\algA(p)$ representations have a natural expression in terms of
Gordon--type formulas.

In Sec.~\ref{short} we recall some known facts about $(1,p)$
logarithmic conformal field models and in Sec.~\ref{theproof} give the
proof of Thm.~\ref{thm:main}.

%  \begin{equation}\label{eq:characters}
%    \begin{aligned}
%      \chi(\repX^+_{s,p})(q)&=\mfrac{1}{\eta(q)}
%      \Bigl(\ffrac{s}{p}\,\theta_{p{-}s,p}(q)
%      + 2\,\theta'_{p{-}s,p}(q)\Bigr),\\[4pt]
%      \chi(\repX^-_{s,p})(q)&=
%      \mfrac{1}{\eta(q)}
%      \Bigl(\ffrac{s}{p}\,\theta_{s,p}(q) - 2\,\theta'_{s,p}(q)\Bigr),
%    \end{aligned}\qquad
%    1\leq s\leq p.
%  \end{equation}
%\begin{align*}
%  \eta(q)&=q^{\frac{1}{24}} \prod_{n=1}^{\infty} (1-q^n)
%  \\
%  \intertext{and the theta functions}
%  \theta_{s,p}(q,z)&=\sum_{j\in\oZ + \frac{s}{2p}} q^{p j^2} z^j,
%  \quad |q|<1,~z\in\oC\,,
%\end{align*}
%and set $\theta_{s,p}(q)\,{:=}\,\theta_{s,p}(q,1)$ and
%$\theta'_{s,p}(q)\,{:=}\,z\frac{\dd}{\dd
%  z}\theta_{s,p}(q,z)\!\!\bigm|_{z=1}$.

\section{Short description of logarithmic $(1,p)$ models\label{short}}

\subsection{Notations}
Throughout the paper we use the standard notation
\begin{equation}
  \alpha_+=\sqrt{2p}\,,\qquad\alpha_-=-\sqrt{\frac{2}{p}}\,,\qquad \alpha_+\alpha_-=-2,
  \qquad \alpha_0=\alpha_++\alpha_-=\sqrt{\frac{2}{p}}(p-1),
\end{equation}
where $p$ is a positive integer.
We let $\repM_{r,s;p}$ denote the irreducible module with the highest weight
\begin{equation}\label{d-conf}
  \Delta_{r,s} =
  \frac{p}{4}(r^2-1) + \frac{1}{4p}(s^2-1) +
  \frac{1 - rs}{2},~~1\leq s\leq p,~r\in\oZ
\end{equation}
of the Virasoro algebra with the central charge~\eqref{c-charge}.  We
note that $\repM_{r,s;p}$ is the quotient of the Verma module by the
submodule generated from one singular vector on the level $rs$ and
such modules exhaust irreducible Virasoro modules that aren't Verma
modules.

In terms of the free scalar field $\varphi$ with the OPE
$\varphi(z)\,\varphi(w) = \log(z{-}w)$ the highest weight vector of
$\repM_{r,s;p}$ corresponds to the vertex field~\cite{[DMS]}
\begin{equation}
  V_{r,s}=\eE^{-(\frac{r-1}{2}\alpha_++\frac{s-1}{2}\alpha_-)\,\varphi(z)}.
\end{equation}
The generators of the Virasoro algebra are Laurent coefficients of the
energy--momentum tensor
\begin{equation}\label{the-emt}
  T=\frac{1}{2}\, \no{\dd\varphi\,\dd\varphi}
  + \frac{\alpha_0}{2}\, \dd^2\varphi.
\end{equation}

\subsection{The triplet $W$-algebra $\algW(p)$}
The triplet $W$-algebra  can be described in terms of the
lattice vertex operator algebra generated by the vertex operators~\cite{[FFHST]}
\begin{equation}
  V^\pm(z)=e^{\pm\alpha_+\varphi(z)}.
\end{equation}
The algebra $\algW(p)$ is a subalgebra of this lattice vertex operator algebra.
%hence
%\begin{equation}
%  \dd\varphi(z)\,\dd\varphi(w) = \zw2 \,.  \label{3a}
%\end{equation}
%\begin{equation}
%\varphi(z)=p_0+h_0\log z-\sum_{i\neq0}\frac{h_i}{i}z^{-i}\,,\qquad
%[h_i,h_j]=i\delta_{i+j,0}\,,\quad [h_0,p_0]=1
%\end{equation}
The vacuum representation of $\algW(p)$ is the kernel of the screening
operator
\begin{equation}
F = \Frac{1}{2\pi\ii}\oint dz e^{\alpha_-\varphi(z)}
\end{equation}
acting in the vacuum representation of the lattice VOA.  This kernel
is generated by the $s\ell(2)$-algebra triplet
\begin{equation}\label{w-gen}
  W^-=e^{-\alpha_+\varphi(z)},~W^0=[e,W^-],~W^+=[e,W^0],
\end{equation}
where
\begin{equation}
  e= \Frac{1}{2\pi\ii}\oint dz e^{\alpha_+\varphi(z)}
\end{equation}
is one of the $s\ell(2)$ algebra generators. The generator $f$ in
terms of $\varphi$ is given by a nonlocal expression.  $\algW(p)$
contains the energy--momentum tensor~\eqref{the-emt} with the central
charge~\eqref{c-charge}.  The generators $W^{\pm,0}$ are primary
fields of dimension $2p-1$.

\subsection{Irreducible representations of~$\algW(p)$}
Each irreducible $\algW(p)$ modules $\repX^\pm_{s,p}$ can be described
in terms of irreducible Virasoro modules~$\repM_{r,s;p}$.  Let $\pi_r$
denote the $r$-dimensional irreducible representation of $s\ell(2)$.
Then the spaces
\begin{align}\label{Xpm-decomp}
  \repX^+_{s,p}=&\oplus_{n\in\oN}\pi_{2n-1}\tensor\repM_{2n-1,s;p},\\
  \repX^-_{s,p}=&\oplus_{n\in\oN}\pi_{2n}\tensor\repM_{2n,s;p}
\end{align}
admit an action of $\algW(p)$ and are its irreducible modules. These decompositions
give formulas~\eqref{chi-pl} and~\eqref{chi-min} for the characters.

\subsection{The algebra $\algA(p)$}
We consider the ``nonlocal'' vertex--operator algebra $\algA(p)$
generated by the $s\ell(2)$ doublet of fields
\begin{equation}\label{defa}
  a^+(z)=e^{-\frac{\alpha_+}{2}\varphi(z)}\,,\qquad
  a^-(z)=[e,a^+(z)]=D_{p-1}(\dd\varphi(z))e^{\frac{\alpha_+}{2}\varphi(z)},
\end{equation}
where $D_{p-1}$ is a degree $p-1$ differential polynomial
in~$\dd\varphi(z)$. The conformal dimension of these fields is
$\frac{3p-2}{4}$.  The fields $a^\pm(z)$ have the following OPE
\begin{equation}\label{mainOPE}
  a^+(z)a^-(w)=(z-w)^{-\frac{3p-2}{2}}\sum_{n\geq0}
  (z-w)^{n}H^n(w)
%+
%  T^2(z-w)^{-\frac{3p-2}{2}+4}+\dots\\
%  +T^{p-1}(z-w)^{\frac{p-2}{2}}+T^{p}(z-w)^{\frac{p-2}{2}+1}+\dots\Bigr),
\end{equation}
where $H^n(w)$ are fields with conformal dimension equals to~$n$.  The field
$H^0$ is proportional to the identity field $1$, $H^1=0$, $H^2$ is proportional to
the energy--momentum tensor~$T$. About other fields $H^n$ we can say the following
\begin{gather}
  \label{H2n}
  H^{2n}=c_{2n} :T^n:+P_{2n}(T),\qquad1\leq n\leq p-1,\\
  H^{2n+1}= c_{2n+1}\partial:T^n:+P_{2n+1}(T),\qquad1\leq n\leq p-2,\\
  H^{2p-1}=c_{2p-1}\partial:T^{p-1}:+P_{2p-1}(T)+d_1W^0,\\
  \label{H2p}
  H^{2p}=c_{2p}:T^p:+P_{2p}(T)+d_2\partial W^0,
\end{gather}
where $:T^n:$ is the normal ordered $n$-th power of the
energy--momentum tensor, $P_{n}(T)$ is a differential polynomial in
$T$ and degree of both $P_{2n}(T)$ and $P_{2n+1}(T)$ in~$T$ is equal
to $n-1$ , $W^0$ is the field defined in~\eqref{w-gen} and $c_n$,
$d_1$, $d_2$ are some nonzero constants.

\subsection{The irreducible representations of $\algA(p)$}
The vertex operator algebra $\algA(p)$ is graded (by eigenvalues of
the zero mode of $\partial\varphi$)
\begin{equation}
  \algA(p)=\bigoplus_{\beta\in\frac{\alpha_+}{2}\oZ}\algA(p)^\beta
\end{equation}
and $a^{\pm}(z)\in\algA(p)^{\pm\frac{\alpha_+}{2}}$.  We consider only
the graded representations of~$\algA(p)$.  For any representation
$\repX=\oplus_{t\in\oC}\repX^t$ we have
$a^{\pm}(z):\repX^t\to\repX^{t\pm\frac{\alpha_+}{2}}$ and $a^{\pm}(z)$
acting in $\repX^t$ have the decomposition
\begin{equation}
\label{modes}
  a^{\pm}(z)=\sum_{n\in\pm t\frac{\alpha_+}{2}-\frac{3p-2}{4}+\oZ}
                     z^{-n-\frac{3p-2}{4}}a^{\pm}_n.
\end{equation}
The irreducible representations $\repX_{s,p}$ of $\algA(p)$ are
highest-weight modules generated from the vector
$\ket{s,p}\in\repX_{s,p}^{\frac{1-s}{2}\alpha_-}$ satisfying
\begin{equation}\label{cond}
  a^\pm_{-\frac{3p-2s}{4}+n}\ket{s,p}=0,\qquad n\in\oN,~
%  H^1_{\leq-1},H^2_{\leq-2},\dots,H^{s-1}_{\leq-(s-1)},
%  H^{s}_{\leq-(s+1)},H^{s+1}_{\leq-(s+2)},\dots,H^{p-1}_{\leq-p},
1\leq s\leq p.
\end{equation}
The conformal dimension of $\ket{s,p}$ is
$\Delta_{1,s}=\frac{s^2-1}{4p}+\frac{1-s}{2}$.  The highest mode of
$a^\pm(z)$ that generate new vectors from $\ket{s,p}$ are
\begin{equation}
  a^\pm_{-\frac{3p-2s}{4}},~1\leq s\leq p
\end{equation}
as it shown
\begin{figure}[tb]
  \mbox{}~\quad~
  \xymatrix@=6pt{%
    &&&&&&&&\\
   &&&&&&&{\bullet}
    \ar@{}[-1,0]|(.8){}
    \ar[3,-2]_{a^-_{\frac{2s-3p}{4}}}
    \ar[3,2]^{a^+_{\frac{2s-3p}{4}}}\\
    &&&&&&&&\\
    &&&&&&&&\\
    &&&&&{\circ}\ar[5,-2]_{a^-_{\frac{2s-5p}{4}}}
    \ar@{}[-1,-2]|(.6){\displaystyle V_{2,s}}
    &&&&{\circ}\ar[5,2]^{a^+_{\frac{2s-5p}{4}}}\\
    &&&&&&&&&&&&\\
    &&&&&&&&&&&&&\\
    &&&&&&&&&&&&\\
    &&&&&&&&&&&&&\\
    &&&{\bullet}\ar[7,-2]_{a^-_{\frac{2s-7p}{4}}}
   \ar@{}[-1,-2]|(.7){\displaystyle V_{3,s}}
    &&&&&&&&{\bullet}\ar[7,2]^{a^+_{\frac{2s-7p}{4}}}\\
    &&&&&&&&&&&&&&&&\\
    &&&&&&&&&&&&&&&&\\
    &&&&&&&&&&&&&&&&\\
    &&&&&&&&&&&&&&&&\\
    &&&&&&&&&&&&&&&&\\
    &&&&&&&&&&&&&&&&\\
    &{\circ}\ar@{.}[];[]+<-3pt,-18pt>
    \ar@{}[0,-1]-<20pt,0pt>|(.6){\displaystyle V_{4,s}}
    &&&&&&&&&&&&{\circ}
    \ar@{.}[];[]+<3pt,-18pt>
    }
  \caption[The  irreducible $\algA(p)$ modules]
  {\small%\captionfont
    {The  irreducible $\algA(p)$ modules.}  The filled dot on the top is the cyclic vector
    $\ket{s,p}$. The arrows show the action of highest modes of
    $a^\pm$ that give nonzero vectors.  Filled (open) dots denote
    vertices belonging to representations $\repX^+_s$ ($\repX^-_s$).}
 \label{fig:LambdaPi}
\end{figure}
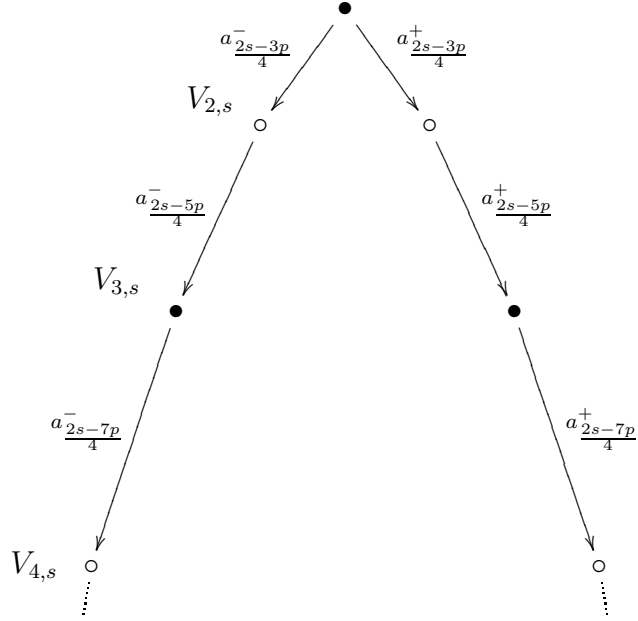
in Fig.~\ref{fig:LambdaPi}.
Proceeding further we obtain the set of extremal vectors shown in Fig.~\ref{fig:LambdaPi}.

From~\eqref{Xpm-decomp}, we immediately obtain that the irreducible
representation~$\repX_{s,p}$ as a representation of
$s\ell(2)\oplus\mathrm{Vir}$ decomposes as
\begin{equation}\label{decomp}
  \repX_{s,p}=\oplus_{n\in\oN}\pi_{n}\tensor\repM_{n,s;p}.
\end{equation}

\begin{rem}
In the rest of the paper we use the notation $\ch V$ for the normailzed
character of the space $V$. Namely the character
$\ch V$ is a Laurent series  $\sum_{i\in\oZ} a_iq^i$ such that
$a_i=0$ for $i<0$ and $a_0\ne 0$. For example for $V=\repX_{s,p}$ we have
$$\chi_{s,p}(q)=
q^{\frac{s^2-1}{4p}+\frac{1-s}{2}-\frac{c}{24}} \ch \repX_{s,p}.$$
The normalization above is natural for us because of the fermionic
(particle) approach used in the paper.
\end{rem}

\section{Proof of Theorem~\ref{thm:main}\label{theproof}}
The strategy of the proof is as follows. We introduce a certain
filtrations on the algebra $\algA(p)$ such that the adjoint graded
algebra $\bar\algA(p)$ can be described in terms of generators and
quadratic relations. We study highest weight representations of
$\bar\algA(p)$ and derive fermionic formula for their characters. We
show that these characters are equal to~$\chi_{s,p}(q)$.

\subsection{Filtrations and adjoint graded algebras}
We introduce a filtration $F_\bullet$ on $\algA(p)$ by attaching
\begin{itemize}
\label{F}
\item
degree $p$ to
each mode of $a^+(z)$,
\item degree $p-1$ to each mode of $a^-(z)$,
\item degree $2$ to each mode of $T(z)$.
\end{itemize}
We denote the adjoint graded algebra with respect to $F_\bullet$ by
$\bar\algA(p)$ and  its generators by $\bar a^\pm(z)$ and~$\bar
T(z)$.

\begin{Lemma}
\label{barA}
The following relations hold in $\bar\algA(p)$:
\begin{gather}
\label{apm}
\bar a^+(z)\bar a^+(w)\sim (z-w)^{\frac{p}{2}},\
\bar a^-(z)\bar a^-(w)\sim (z-w)^{\frac{p}{2}},\\
\label{Ta}
\bar T(z)\bar a^\pm(w)\sim z-w,\\
\label{a+a-}
\bar a^+(z)\bar a^-(w)\sim (z-w)^{\frac{p}{2}},
\end{gather}
where $A(z)B(w)\sim(z-w)^x$ means that the fields $A(z)$ and $B(z)$
have the following OPE
\begin{equation}
  A(z)B(w)=(z-w)^x\sum_{n\geq0}(z-w)^nC^n(w)
\end{equation}
with some fields~$C(z)$.  In addition the current $\bar T(z)$ is
commutative and satisfy $\bar T(z)^p=0$.
\end{Lemma}
\begin{proof}
From the formula \eqref{defa} we obatin that in $\algA(p)$ the following
is true:
$$\bar a^+(z)\bar a^+(w)\sim (z-w)^{\frac{p}{2}}.$$
Therefore the first part of $(\ref{apm})$ holds in $\algA(p)$.
The second part of $(\ref{apm})$ follows from the first part and an equation
$[e,a^-(z)]=0$ (see \cite{[FGST2]}).

To prove $(\ref{Ta})$ we use the OPE in $\algA(p)$:
$$T(z)a^\pm(w)=(z-w)^{-1} a^{\pm}(w)+:T(z) a^\pm(w):+\dots.$$
Now using the relation
$$
:T(z)a^\pm(z):+\mathrm{const}\cdot \partial^2 a^\pm(z)=0
$$
(recall that $a^\pm(z)$ are two components of the field $\Phi_{2,1}(z)$)
we obtain  $(\ref{Ta})$.
We now prove $(\ref{a+a-})$.

The OPE \eqref{mainOPE} gives the following OPE in $\bar\algA(p)$:
\begin{multline}
\label{barOPE}
\bar a^+(z)\bar a^-(w)=(z-w)^{-\frac{3p-2}{2}}
\Bigl[\sum_{i=0}^{2(p-1)} (z-w)^i \bar H_i(z)) + \\
(z-w)^{2p-1} (c_{2p-1} \partial :\bar T^{p-1}(z): + d_1 \bar W^0(z))+
(z-w)^{2p}(c_{2p} :\bar T^p (z): + d_2 \partial\bar W^0(z))\Bigr]+
\dots
\end{multline}
(see $(\ref{H2n})-(\ref{H2p})$).
We recall that for any $0\le i\le 2p-2$ the operator $H_i(z)$ is a differential
polynomial in $T(z)$ of degree smaller than or equal to $p-1$. Therefore
the degree of $\bar H_i$ with respect to our filtration is smaller than
$2p-1$. But the degree of $\bar a^+(z)\bar a^-(w)$ (which is the left hand
side of ($\ref{barOPE}$)) is exactly $2p-1$. Therefore we can rewrite
$(\ref{barOPE})$ as
\begin{equation}
\label{W0}
\bar a^+(z)\bar a^-(w)=(z-w)^{-\frac{3p-2}{2}}
\Bigl[(z-w)^{2p-1} d_1\bar W^0(z)+
(z-w)^{2p}(c_{2p} :\bar T^p (z): + d_2 \partial\bar W^0(z))\Bigr]+\dots.
\end{equation}
This proves $(\ref{a+a-})$.

We now consider the current $\bar T(z)$. Note that $\bar T(z)$ is
commutative, because the degree of each mode of $\bar T(z)$ is equal to $2$.
We now show that $\bar T^p=0$. From $(\ref{W0})$ we obtain that  each
mode of the current $\bar W^0(z)$ can be
expressed as a linear combination of $\bar a^+_i\bar a^-_j$. The same is
true for the modes of
$c_{2p} :\bar T^p (z): + d_2 \partial\bar W^0(z)$ and thus for
$\bar T^p (z)$. But the degree of $\bar T^p(z)$ equals $2p$ and the degree of
$\bar a^+_i\bar a^-_j$ is equal to $2p-1$. This gives $\bar T^p(z)=0$.
Lemma is proved.
\end{proof}

We now want to replace the condition $\bar T^p(z)=0$ by the set of quadratic
relations. We use the standard Lemma (see \cite{[FS],[FJKLM]}).

\begin{Lemma}
\label{T^p}
Let $\algB$ be the algebra generated by modes $J_0, J_1,\dots$ of an abelian
current $J(z)$ with the
defining relation $J(z)^p=0$.
There exists a filtration  $G_\bullet$
on the algebra $\algB$ such that the adjoint graded algebra is
generated by coefficients of series $J^{[i]}(z)$, which are images of
powers $J(z)^i$, $1\le i <p$.
In addition defining relations in the adjoint graded algebra
are given by
\begin{equation}
\label{Ti}
J^{[n]}(z) J^{[m]}(w)\sim (z-w)^{2{\mathrm min}(n,m)}.
\end{equation}
\end{Lemma}

We now consider the  filtration on  $\bar\algA(p)$ induced from the filtration
$G_\bullet$ on the algebra generated with the modes of $\bar T(z)$. We
denote the adjoint graded algebra by the same symbol $\bar\algA(p)$. In what
follows we use the notation $\bar\algA(p)$ to denote the adjoint graded
algebra of $\algA(p)$ with respect to the double filtration ($F_\bullet$
and $G_\bullet$).

We now introduce a new algebra which is quadratic with the defining relations
given by $(\ref{Ti})$, $(\ref{apm})$, $(\ref{Ta})$ and $(\ref{a+a-})$.
\begin{Dfn}
Let $\bar\algA(p)'$ denote an algebra generated with the currents
$$\bar a^+(z), \bar a^-(z), \bar T^{[i]}(z), 1\le i <p$$
and defining relations
\begin{gather}
\label{p/2}
\bar a^\pm(z)\bar a^\pm(w)\sim (z-w)^{\frac{p}{2}},\\
\label{aTn}
 \bar a^\pm(z) \bar T^{[n]}(w)\sim (z-w)^n,\\
\label{TT}
\bar T^{[n]}(z) \bar T^{[m]}(w)\sim (z-w)^{2{\mathrm min}(n,m)}.
\end{gather}
\end{Dfn}

We note that Lemmas $\ref{barA}$ and $\ref{T^p}$ gives a surjection
\begin{equation}
\label{AtoA}
\bar\algA(p)'\to \bar\algA(p).
\end{equation}
We define the $s\ell(2)$ action on $\bar\algA(p)'$ as follows.  $\bar
T^{[n]}(z)$ are $s\ell(2)$ invariants and $\bar a^+(z)$ and $\bar a^-(z)$
are highest and lowest weight vectors of the $s\ell(2)$ doublet
respectively. This action commutes with the
mapping~\eqref{AtoA}.

We now study highest weight representations of $\bar\algA(p)'$.  Let
$\bar \repX_{s,p}'$, $1\leq s\leq p$ denote the cyclic representation
of $\bar\algA(p)'$ that is generated from the vector $v_{s,p}$
satisfying the defining relations (The Fourier decomposition of $\bar
a^\pm(z)$ is the same as in~\eqref{modes}.)
\begin{gather}
  \label{3p-2s}
  \bar a^\pm_j v_{s,p}=0,\quad j>-\frac{3p-2s}{4},\\
  \label{[n]}
   \bar T^{[n]}_jv_{s,p}=0,\quad j>\left\{
    \begin{array}{l}
     -n,\quad n<s,\\
     -2n+s-1,\quad n\geq s.
    \end{array}
\right.
\end{gather}

We note that because of $(\ref{AtoA})$ and $(\ref{cond})$ there exists
a surjective homomorphism
\begin{equation}
\label{XtoX}
\bar \repX_{s,p}'\to \bar\repX_{s,p},
\end{equation}
where $\bar\repX_{s,p}$ is an adjoint graded to $\repX_{s,p}$ with respect to
the filtrations induced from $F_\bullet$ and $G_\bullet$. In particular,
the characters of
$\bar \repX_{s,p}$ and $\repX_{s,p}$ coincide and
$\bar \repX_{s,p}\simeq \repX_{s,p}$ as $s\ell(2)$ modules.

\begin{Lemma}
\label{chx'}
The character of $\bar \repX_{s,p}'$ is given by the right hand side of the
formula  $(\ref{eq:ferm-char})$.
\end{Lemma}

\begin{proof}
We briefly recall the functional realization of the dual space (see for
example ~\cite{[FLT]}).

Consider the decomposition
$$
\bar\repX_{s,p}'=\bigoplus_{n_+,n_-,n_1,\dots, n_{p-1}\ge 0}
\bar\repX_{s,p}'(n_+,n_-,n_1,\dots, n_{p-1}),
$$
where $\bar\repX_{s,p}'(n_+,n_-,n_1,\dots, n_{p-1})$
 is the linear span of the vectors of the form
$$
\left[\bar a^+_{i^+_1}\dots\bar a^+_{i^+_{n_+}} \bar a^-_{i^-_1}\dots\bar a^-_{i_{n_-}}
\prod_{\alpha=1}^{p-1} \bar T^{[\alpha]}_{i^\alpha_1}\dots
\bar T^{[\alpha]}_{i^\alpha_{n_\alpha}}\right]\cdot  v_{s,p}
$$
with arbitrary parameters $i^{\pm}_\beta, i^\alpha_\beta$.
For $\theta\in (\bar\repX_{s,p}'(n_+,n_-,n_1,\dots, n_{p-1}))^*$ we consider a
correlation function
\begin{multline*}
F_\theta=\langle \theta| \bar T^{[1]}(x_1^1)\dots \bar T^{[1]}(x_{n_1}^1)
\dots \bar T^{[p-1]}(x_1^{p-1})\dots \bar T^{[p-1]}(x_{n_{p-1}}^{p-1})\\
\bar a^+(x_1^+)\dots \bar a^+(x_{n_+}^+)
\dots \bar a^-(x_1^-)\dots \bar a^-(x_{n_-}^-)| v_{s,p}\rangle.
\end{multline*}
The space of thus obtained functions can be described as the space of
functions of the form
$$(f\cdot X) (x_1^1, \dots, x_{n_1}^1, \dots, x_1^{p-1}, \dots, x_{n_{p-1}}^{p-1},
x_1^+, \dots, x_{n_+}^+, x_1^-, \dots, x_{n_-}^-),$$
where $X$ is a function given by the formula
\begin{multline}
\label{dualspace}
(\prod_{1\le i\le n_+} x^+_i \prod_{1\le j\le n_- } x^-_j)^{\frac{3p-2s}{4}}
\prod_{1\le \al\le s-1} (\prod_{1\le i\le n_\al} x^{\al}_i)^\al
\prod_{s\le \al\le p-1} (\prod_{1\le i\le n_\al} x^{\al}_i)^{2\al-s+1}\\
\prod_{a=+,-}\prod_{1\le i\le n_a} (x^a_i-x^b_j)^{p/2}
\prod_{\atop{1\le\al \le p-1}{1\le i< j\le n_{\al}}}
(x^{\al}_i-x^{\al}_j)^{2\al}\\
\prod_{\atop{1\le i\le n_a}{1\le j\le n_b}} (x^+_i-x^-_j)^{p/2}
\prod_{\atop{b=+,-}{1\le \al\le p-1}}
\prod_{\atop{1\le i\le n_a}{1\le j\le n_{\al}}}
(x^b_i-x^\al_j)^\al
\prod_{1\le\al<\be \le p-1}
\prod_{\atop{1\le i\le n_{\al}}{1\le j\le n_{\be}}}
(x^{\al}_i-x^{\be}_j)^{2\min(\al,\be)},
\end{multline}
and $f$ is a polynomial symmetric in each group of variables
$$\{x^+_i\}_{i=1}^{n_+},\ \{x^-_i\}_{i=1}^{n_-},\ \{x^{\al}_i\}_{i=1}^{n_\al},
\al=1,\dots,s.$$
The exact form $(\ref{dualspace})$ of the functions $F_\theta$ follows from the
definition of $\bar\algA(p)'$ as an algebra with defining quadratic relations
and from the definition of $\bar\repX'_{s,p}$.
In particular the factor in the first line of $(\ref{dualspace})$ comes
from the relations  $(\ref{3p-2s})$ , $(\ref{[n]})$ and the the rest factors
correspond to  $(\ref{p/2})$, $(\ref{aTn})$, $(\ref{TT})$.
Direct computation shows that the character of the space of polynomials
$(\ref{dualspace})$ is given by the
right
hand side of the formula $(\ref{eq:ferm-char})$.
\end{proof}

This Lemma gives an upper bound for the character of $\repX_{s,p}$. To prove
that~\eqref{XtoX} is an isomorphism, we consider the decomposition
  \begin{equation}
    \bar \repX_{s,p}'=\bigoplus_{n=1}^\infty \pi_n\tensor \bar \repX_{s,p}'[n],
  \end{equation}
where $\bar \repX'_{s,p}[n]$ is a space of multiplicity of $\pi_n$ in
$\bar \repX_{s,p}$.
Our goal is to show that
  \begin{equation}
  \label{main}
    \mbox{\rm ch}\bar \repX_{s,p}'[r]=\mbox{\rm ch}\repM_{r,s;p}
  \end{equation}
Because of the surjection $(\ref{XtoX})$ and formula $(\ref{decomp})$
the proof of the equation $(\ref{main})$ is enough for the proof of the
Theorem $\ref{thm:main}$.

We divide the proof of $(\ref{main})$ into 2 parts:
we first show that
\begin{equation*}
      \mbox{\rm ch}\bar\repX'_{s,p}[1]=\mbox{\rm ch}\repM_{1,s;p}
  \end{equation*}
and then deduce the general $r$ case.

\subsection{The proof of
$\mbox{\rm ch}\bar\repX'_{s,p}[1]=\mbox{\rm ch}\repM_{1,s;p}$}
We first let $p=1$ and consider the decomposition
\begin{equation}
\label{Vd}
\bar\repX'_{1,1}=\bigoplus_{n\ge 0} V_n,
\end{equation}
where $V_0$ is spanned by the highest weight vector and
\begin{equation}
V_{n+1}=\mathrm{span}
\langle \bar a_i^+ \bar a_{j_1}^-\dots \bar a_{j_l}^-  v, v\in V_n\rangle,
\end{equation}
with arbitrary $i, j_1,\dots, j_l$. Equivalently,
$$
V_n=\mathrm{span}\langle \bar a^+_{i_1}\dots \bar a^+_{i_n}
\bar a_{j_1}^-\dots \bar a_{j_l}^-  v_{1,1}\rangle,
$$
with arbitrary numbers $i_\al$, $j_\be$ and $l$. We note that the decomposition
$(\ref{Vd})$ is induced from the grading on $\bar\algA(1)'$, which
assignes degree $1$ to each mode of $\bar a^+(z)$ and degree $0$ to each mode
of $\bar a^-(z)$. We note also that this construction applied to the
algebra $\algA(1)$ produces exactly the filtration $F_\bullet$
(see the beginning of the subsection $\ref{F}$).

For any $M$ with an action of an operator $h$ and $l\in\mathbb{Z}$ we set
$$M^l=\{v\in M:\ hv=lv\}$$
($h$ is a standard generator of the Cartan subalgebra of $s\ell(2)$).

\begin{Lemma}
Let $l\ge 0$. If $l>n$ then $V^l_n=0$. Otherwise
\begin{equation}
\ch V_n^l=\frac{q^{(n-\frac{l}{2})^2}}{(q)_n (q)_{n-l}}.
\end{equation}
\end{Lemma}
\begin{proof}
We recall that $\bar a^+(z)$ and $\bar a^-(z)$ for two-dimensional
irreducible representation of $s\ell(2)$. Therefore $V_n^l$ is the linear span
of a set of vectors
$$\bar a^+_{i_1}\dots \bar a^+_{i_n} \bar a^-_{j_1}\dots \bar a^-_{j_{n-l}}
v_{1,1}$$
with arbitrary $i_\al$, $j_\be$. This leads to the description of the dual
space $(V^l_n)^*$ as the space of polynomials in variables
$x^+_1,\dots, x^+_n$, $x^-_1,\dots, x^-_{n-l}$ of the form
\begin{equation*}
(\prod_{i=1}^n x^+_i \prod_{j=1}^{n-l} x^-_j)^{\frac{1}{4}}
\bigl[\prod_{1\le i< j\le n} (x^+_i-x^+_j)
\prod_{1\le i< j\le n-l} (x^-_i-x^-_j)
\prod_{\atop{1\le i\le n}{1\le j\le n-l}} (x^+_i-x^-_j)\bigr]^{\frac{1}{2}}
\times g,
\end{equation*}
where $g(x^+_1,\dots, x^+_n, x^-_1,\dots, x^-_{n-l})$ is an arbitrary
polynomial symmetric in each group of variables
$\{x^+_i\}_{i=1}^n$ and $\{x^-_j\}_{j=1}^{n-l}$.
The degree of the product above is equal to
$(n-l/2)^2+\deg g$. Lemma is proved.
\end{proof}

Set $V_n[1]=V_n\cap \bar\repX'_{1,1}[1]$.

\begin{Prop}
\label{todaf}
\begin{equation}
\label{toda}
\ch V_n[1]=
\sum_{\atop{n_1,n_2,\dots\ge 0}{\sum n_i=n}}
\frac{q^{\frac{1}{2}\sum_{i,j\ge 1} 2\min(i,j) n_in_j+\sum_{i\ge 1} in_i   }}
{(q)_{n_1}(q)_{n_2}\dots}.
\end{equation}
\end{Prop}

\begin{proof}
Using the relation $\ch V_n[1]=\ch V_n^0-\ch V_{n+1}^2$ and Lemma above
we obtain
$$\ch V_n[1]=\frac{q^{n^2}}{(q)_n^2}-\frac{q^{n^2}}{(q)_{n+1}(q)_{n-1}}=
\frac{q^{n^2}q^n(1-q)}{(q)_n(q)_{n+1}}.$$
So we need to show that
\begin{equation}
\label{bf}
\frac{q^{n^2}q^n(1-q)}{(q)_n(q)_{n+1}}=
\sum_{\atop{n_1,n_2,\dots\ge 0}{\sum n_i=n}}
\frac{q^{\frac{1}{2}\sum_{i,j\ge 1} 2\min(i,j) n_in_j+\sum_{i\ge 1} in_i   }}
{(q)_{n_1}(q)_{n_2}\dots}.
\end{equation}
Instead we prove a more general relation
\begin{equation}
\label{ubf}
\frac{q^{n^2}u^n}{(q)_n(uq)_n}=
\sum_{\atop{n_1,n_2,\dots\ge 0}{\sum n_i=n}}
\frac{q^{\frac{1}{2}\sum_{i,j\ge 1} 2\min(i,j) n_in_j}
u^{\sum_{i\ge 1} in_i}}
{(q)_{n_1}(q)_{n_2}\dots}.
\end{equation}
where a new variable $u$ is introduced and the notation
$(uq)_n=(1-uq)(1-uq^2)\dots (1-uq^n)$ is used. We note that the relation above
reduces to $(\ref{bf})$ after the specialization $u=q$.

After the change of varibales $m_i=n_{i+1}+n_{i+1}+\dots$, $i=1,2,\dots$
the equation $(\ref{ubf})$ becomes
\begin{equation}
\frac{q^{n^2}u^n}{(q)_n(uq)_n}=
\sum_{n\ge m_1\ge m_2\ge  \dots\ge 0}
\frac{q^{\sum_{i\ge 1} m_i^2} u^{\sum_{i\ge 1} m_i} q^{n^2} u^n}
{(q)_{n-m_1}(q)_{m_1-m_2}\dots},
\end{equation}
or equivalently
\begin{equation}
\label{binom}
\frac{1}{(uq)_n}=
\sum_{n\ge m_1\ge m_2\ge \dots\ge 0}
q^{\sum_{i\ge 1} m_i^2} u^{\sum_{i\ge 1} m_i}
\binom{n}{m_1}_q \binom{m_1}{m_2}_q\dots,
\end{equation}
where the notation $\binom{m}{n}_q=\frac{(q)_n}{(q)_m(q)_{n-m}}$ is used
for a $q$-binomial coefficient.

We prove $(\ref{binom})$ by induction on $n$. The case $n=1$ is obvious.
For general $n$ we rewrite the right hand side of $(\ref{binom})$ as
\begin{equation}
\sum_{m_1=0}^n q^{m_1^2}  \binom{n}{m_1}_q u^{m_1}
\sum_{m_1\ge m_2\ge \dots\ge 0}
q^{\sum_{i\ge 2} m_i^2} u^{\sum_{i\ge 2} m_i}
\binom{m_1}{m_2}_q \binom{m_2}{m_3}_q\dots.
\end{equation}
Therefore using the induction assumption it is enough to show that
\begin{equation}
\label{frac}
\frac{1}{(uq)_n}=
\sum_{m=0}^n q^{m^2} u^m \binom{n}{m}_q \frac{1}{(uq)_m}.
\end{equation}
The left hand side is equal to the $(u,q)$ character of the space of
polynomials in commuting variables $e_i$, $1\le i\le n$, where $\deg _u e_i=1$ and
$\deg_q e_i=i$. We consider the decomposition
$$
\mathbb{C}[e_1,\dots, e_n]=\mathbb{C}\cdot 1 \oplus
\bigoplus_{m=1}^n \mathbb{C}[e_1, \dots, e_m]\cdot
\mathrm{span}\langle e_{i_1}\dots e_{i_m},\
m\le i_1\le \dots\le i_m\le n \rangle.
$$
The $(u,q)$ character of the right hand side is equal to the right hand side
of $(\ref{frac})$. This finishes the proof of the proposition.
\end{proof}

Consider the space
$\bar\repX'^a_{s,p}\hookrightarrow \bar\repX'_{s,p}$, which is generated
from the
highest weight vector with the modes of $\bar a^{\pm}(z)$ (but not
$\bar T^{[i]}(z)$). We have a decomposition
$$\bar\repX'^a_{s,p}=
\bigoplus_{r\ge 1} \pi_r\otimes \bar\repX'^a_{s,p}[r].$$

\begin{Lemma}
\label{dualsp}
For any $p$ the dual space  $(\bar\repX'^a_{s,p}[1])^*$  is isomorphic to the
direct
sum over $n\ge 0$ of spaces of functions of the form
\begin{equation}
\label{ds}
  (x_1\dots x_{2n})^{\frac{3p-2s}{4}}
   \prod\limits_{1\leq i<j\leq 2n}  (x_i-x_j)^{p/2}
 g(x_1,\dots, x_{2n}),
\end{equation}
where $g(x_1,\dots,x_{2n})$ is a polynomial with values in
the space $(\pi_2^{\tensor 2n})^{s\ell(2)}$, which satisfy
\begin{equation}
\label{sigma}
\sigma_{i,j} g(\dots, x_j,\dots, x_i,\dots)= g(\dots, x_i,\dots, x_j,\dots),
\end{equation}
where $\sigma_{i,j}$ is a transposition acting on $\pi_2^{\tensor 2n}$ by
permuting $i$-th and $j$-th factors.
\end{Lemma}
\begin{proof}
We start with the polynomial realization of the dual space
$(\bar\repX'^a_{s,p})^*$. Let $w_+$, $w_-$ be the standard basis of the
$2$-dimensional irreducible representation of $s\ell(2)$.
For $\theta\in (\bar\repX'^a_{s,p})^*$ we set
$$G_\theta(x_1,\dots,x_k)=\sum_{\al_i=\pm}
\langle \theta | a^{\al_1}(x_1)\dots a^{\al_k}(x_k) | v_{s,p}\rangle
w_{\al_1}\tensor\dots \tensor w_{\al_k}.
$$
This gives a map from $(\bar\repX'^a_{s,p})^*$ to the space of polynomials
in variables $x_1,\dots,x_k$ with values in $\pi_2^{\tensor k}$.
From $(\ref{p/2})$ and $(\ref{3p-2s})$ we obtain that the image of this map
coincides with the subspace $(\ref{ds})$ with restriction $(\ref{sigma})$.

Note that  if
$a^{\al_1}_{i_1}\dots a^{\al_k}_{i_k} v_{s,p}$ belongs to
$\bar\repX'^a_{s,p}[1]$ then the number of pluses and minuses
in the set $\{\al_i\}_{i=1}^k$ coincide. Therefore  to obtain the polynomial
realization of the space $(\bar\repX'^a_{s,p}[1])^*$ one needs to take
an even number $k=2n$ of variables and the space of polynomials with values in
$(\pi_2^{\tensor 2n})^{s\ell(2)}$.
\end{proof}

Consider the decomposition
$$\bar\repX'^a_{s,p}[1]=\bigoplus_{n\ge 0} (\bar\repX'^a_{s,p}[1])_n,$$
where $(\bar\repX'^a_{s,p}[1])_n$ is a subspace defined by the formula
$$\mathrm{span}\langle \bar a^+_{i_1}\dots \bar a^+_{i_n}
\bar a^-_{j_1} \dots \bar a^-_{j_n}\rangle \cap \bar\repX'^a_{s,p}[1]$$
with arbitrary $i_\al, j_\be$. We note that in the case $p=1$
\begin{equation}
\label{p=1}
(\bar\repX'^a_{1,1}[1])_n=V_n[1].
\end{equation}

\begin{cor}
\label{charsp}
\begin{equation}
\label{char}
\ch (\bar\repX'^a_{s,p}[1])_n=
\sum_{\atop{n_p,n_{p+1},\dots\ge 0}{\sum n_i=n}}
\frac{q^{\frac{1}{2}\sum_{i,j\ge p} 2\min(i,j) n_in_j+
\sum_{i\ge p} (i-s+1)n_i}}{(q)_{n_1}(q)_{n_2}\dots}.
\end{equation}
\end{cor}
\begin{proof}
From the Proposition $\ref{todaf}$ and formula $(\ref{p=1})$ we obtain our
Corollary in the case $p=~1$. We now compare the dual space
description $(\ref{ds})$  for general $(s,p)$ and $s=p=1$. The difference of
the degrees is given by the formula
$$
\deg (\prod_{i=1}^{2n} x_i^{\frac{3p-2s}{4}}
   \prod\limits_{1\leq i<j\leq 2n}  (x_i-x_j)^{p/2})-
\deg (\prod_{i=1}^{2n} x_i^{\frac{1}{4}}
   \prod\limits_{1\leq i<j\leq 2n}  (x_i-x_j)^{1/2})=n^2(p-1)+n(p-s).
$$
This gives
\begin{multline*}
\ch (\bar\repX'^a_{s,p}[1])_n= q^{n^2(p-1)+n(p-s)}
\sum_{\atop{n_1,n_2,\dots\ge 0}{\sum n_i=n}}
\frac{q^{\frac{1}{2}\sum_{i,j\ge 1} 2\min(i,j) n_in_j+
\sum_{i\ge 1} in_i}}{(q)_{n_1}(q)_{n_2}\dots}=\\
\sum_{\atop{n_1,n_2,\dots\ge 0}{\sum n_i=n}}
q^{(p-1)(\sum_{i\ge 1} n_i)^2+(p-s)(\sum_{i\ge 1} n_i)}
\frac{q^{\frac{1}{2}\sum_{i,j\ge 1} 2\min(i,j) n_in_j+
\sum_{i\ge p} in_i}}{(q)_{n_1}(q)_{n_2}\dots}.
\end{multline*}
We now redefine $n_i\to n_{p-1+i}$. Then the formula above gives the
right hand side of $(\ref{char})$.
\end{proof}

\begin{Lemma}
\label{Virch}
The character of $\repM_{1,s;p}$ is given by the Gordon type formula
\begin{equation}
  \mbox{\rm ch}\repM_{1,s,p}=\sum_{n_1,n_2,\dots\ge 0}
\frac{q^{\frac{1}{2}\sum_{1\le i\le j} 2\min(i,j)n_in_j
   +n_s+2n_{s+1}+\dots}}{(q)_{n_1}(q)_{n_2}\dots}
\end{equation}
\end{Lemma}
\begin{proof}
We recall that $\repM_{1,s;p}$ is a quotient of the Verma  module $V_{1,s;p}$
by a submodule generated with the a
singular vector on the level $s$
(see \cite{[DMS]}).
Introduce a filtration $H_\bullet$ on
the Verma module $V_{1,s;p}$ defined as follows: $H_0$ is spanned by the highest weight
vector and
$$H_{l+1}=\mathrm{span}\{L_nv,\ v\in H_l, n< 0\}+ H_l.$$
In the corresponding adjoint graded space the images of the operators $L_n$
commute
with each other; we denote these operators as $L_n^{ab}$. This gives
\begin{equation}
\label{quot}
\ch\repM_{1,s;p}=\ch\mathbb{C}[L_{-1}^{ab},L_{-2}^{ab},\dots]/
\{p(L_{-i}^{ab})\},
\end{equation}
where $\{p\}$ is
an ideal generated by some degree $s$ polynomial $p(L^{ab}_{i})$
(we put $\deg L_i^{ab}=-i$). The character of this quotient is independent on
$p(L_{-i}^{ab})$ (only
the degree $s$ matters). We fix $p$ to be equal to $(L_{-1}^{ab})^s$.

Let $T^{ab}(z)=L_{-1}^{ab} +z L_{-2}^{ab}+\dots$. For $k\ge s$ let $R_k$ be a
following ring:
$$R_k=\mathbb{C}[L_{-i}^{ab}]/\{T^{ab}(z)^{k+1}, (L^{ab}_{-1})^s\}.$$
Then for the character of $R_k$ one has a formula
$$\ch R_k= \sum_{n_1,\dots, n_k \ge 0}
\frac{q^{\frac{1}{2}\sum_{1\le i\le j\le k} 2\min(i,j)n_in_j
+\sum_{i\ge s} (s-i+1) n_i}}{(q)_{n_1}\dots (q)_{n_k}}$$
(see \cite{[FKLMM]}).  Obviously
$$
\ch\repM_{1,s;p}=\ch\mathbb{C}[L_{-1}^{ab},L_{-2}^{ab},\dots]/
\{(L_{-1}^{ab})^s\}=\lim_{k\to\infty} \ch R_k.
$$
Lemma is proved.
\end{proof}

\begin{Lemma}
The dual space $(\bar\repX'_{s,p}[1])^*$ is isomorphic to the direct sum
over $n_1,\dots, n_{p-1}, n\ge 0$ of the $(\pi_2^{\otimes 2n})^{s\ell(2)}$
valued polynomials
$$H(x_1^1,\dots, x^1_{n_1},\dots, x^{p-1}_1,\dots, x^{p-1}_{n_{p-1}},
x_1,\dots, x_{2n})$$
of the form $Y\cdot g$, where $Y$ is a function of the form
\begin{multline}
\label{daT}
\prod_{1\le i\le 2n} x_i^{\frac{3p-2s}{4}}
\prod_{1\le i<j \le 2n} (x_i-x_j)^{p/2}
\times\\
\prod_{\atop{1\le i\le 2n}{1\le \al\le p-1}}
\prod_{1\le j\le n_\al} (x_i-x^\al_j)^\al\times\\
\prod_{1\le \al\le s-1} (\prod_{1\le i\le n_\al} x^{\al}_i)^\al
\prod_{s\le \al\le p-1} (\prod_{1\le i\le n_\al} x^{\al}_i)^{2\al-s+1}
\prod_{1\le\al \le p-1} \prod_{1\le i< j\le n_{\al}}
(x^{\al}_i-x^{\al}_j)^{2\al}\\
\prod_{1\le\al<\be \le p-1}
\prod_{\atop{1\le i\le n_{\al}}{1\le j\le n_{\be}}}
(x^{\al}_i-x^{\be}_j)^{2\min(\al,\be)}
\end{multline}
and $g$ is  $(\pi_2^{\otimes 2n})^{s\ell(2)}$ valued
polynomial symmetric in each group of variables
$$(x_1^\al,\dots, x_{n_\al}^\al), \al=1,\dots, p-1;\quad (x_1,\dots,x_{2n}).$$
In addition $g$ satisfies the condition $(\ref{sigma})$ in variables
$x_1,\dots, x_{2n}$.
\end{Lemma}
\begin{proof}
Recall that the currents $\bar T^{[i]}(z)$ commute with the action of
$s\ell(2)$. Therefore we obtain
\begin{equation}
\label{prod}
\bar\repX'_{s,p}[1]=\mathbb{C}[\bar T^{[i]}_j]\cdot (\bar\repX'^a_{s,p}[1]),
\end{equation}
i.e. the space of invariants $\bar\repX'_{s,p}[1]$ can be obtained
by applying all polynomials in modes of the currents  $T^{[i]}(z)$  to
vectors of $\bar\repX'^a_{s,p}[1]$.
The formula $(\ref{daT})$ is a dual version of $(\ref{prod})$. Namely
the first line of $(\ref{daT})$ comes from the polynomial realization of
$(\bar\repX'^a_{s,p}[1])^*$ (see Lemma $\ref{dualsp}$). The second line
of $(\ref{daT})$ describes the interaction between
$\bar a^{\pm}(z)$ and $\bar T^{[i]}(z)$ in the algebra $\bar\algA(p)'$.
Finally the last two lines of $(\ref{daT})$ comes from the polynomial
realization of the dual space of the part of $\bar\repX'_{s,p}[1]$
generated by modes  of $\bar T^{[i]}(z)$.
\end{proof}

\begin{cor}
\label{xsp}
$$
\ch \bar\repX'_{s,p}[1]=
\sum_{n_p,n_{p+1},\dots\ge 0}
\frac{q^{\frac{1}{2}\sum_{i,j\ge 1} 2\min(i,j) n_in_j+
\sum_{i\ge s} (i-s+1)n_i}}{(q)_{n_1}(q)_{n_2}\dots}.
$$
\end{cor}
\begin{proof}
We note that because of the Corollary $\ref{charsp}$ the character
of the space $(\ref{daT})$ is given by a sum over $n_1,\dots,n_{p-1},n\ge 0$
of the terms
\begin{multline*}
\left[ \sum_{n_p+n_{p+1}+\dots=n}
\frac{q^{\frac{1}{2}\sum_{i,j\ge p} 2\min(i,j) n_in_j+
\sum_{i\ge p} (i-s+1)n_i}}{(q)_{n_1}(q)_{n_2}\dots}\right] \times\\
q^{\sum_{\al=1}^{p-1} 2\al n n_\al}\times\\
\frac{q^{\frac{1}{2}\sum_{1\le i,j\le p-1} 2\min(i,j) n_in_j+
\sum_{s\le i\le p-1} (i-s+1)n_i}}{(q)_{n_1}\dots (q)_{n_{p-1}}},
\end{multline*}
where the first line is the character of the first line of $(\ref{daT})$,
the second line is the character of the second line of $(\ref{daT})$
and the third line is the character of the last two lines of $(\ref{daT})$.
Rewriting the formula above in terms of $n_i$, $i>0$ we obtain our Corollary.
\end{proof}

\begin{Prop}
$\ch \bar\repX'_{s,p}[1]= \mbox{\rm ch}\repM_{1,s;p}$.
\end{Prop}
\begin{proof}
Follows from Lemma $\ref{Virch}$ and Corollary $\ref{xsp}$.
\end{proof}

\subsection{The general case}
We recall the decomposition
$$\repX_{s,p}=\bigoplus_{r\ge 1} \pi_r\otimes\repX_{s,p}[r].$$
Our goal is to show that
\begin{equation}
\label{xy}
\repX_{s,p}[r]\simeq \bar\repX'_{s,p}[r].
\end{equation}
We first recall the surjection~\eqref{XtoX}
$$\beta_{s,p}:\bar\repX'_{s,p}\to \repX_{s,p},$$
which is a
homomorphism of $s\ell(2)$ modules. This gives a surjection
$$\beta_{s,p}[r]:\bar\repX'_{s,p}[r]\to \repX_{s,p}[r].$$

Suppose $(\ref{xy})$ doesn't satisfy. Then $\beta_{s,p}$ is not an
embedding.  From the previous section we know that $\beta_{s,p}[1]$ is
an isomorphism.  Denote by $K_r$ the kernel of $\beta_{s,p}[r]$.  Let
$r$ be a minimal number such that $K_r$ is not trivial.  Fix a vector
$u$ which is a highest weight vector of some finite-dimensional
$s\ell(2)$ module $M\simeq \pi_r\in K_r$.  We note that
for any $n\in\oZ$ the space $\langle\bar a^{\pm}_n v\rangle$ spanned by
$\bar a^{\pm}_n v$ with $v\in M$ can be embedded (as $s\ell(2)$
module) to $M\otimes \pi_2$.  In addition $\langle\bar a^{\pm}_n
v\rangle$ is a subspace of
$K_{r+1}\otimes \pi_{r+1}\oplus K_{r-1}\otimes \pi_{r-1}$ because $s\ell(2)$
acts on $\bar a^{\pm}_n$ as on two-dimensional irreducible
representation.  Because of $K_{r-1}=0$, we obtain that for any
$n\in\oZ$
\begin{equation}
\label{r}
r\bar a_n^- u +\bar a^+_n eu=0.
\end{equation}
In fact, the condition $K_{r-1}=0$ means that the linear combination
$\al \bar a_n^- u +\be \bar a^+_n eu$ vanishes whenever
$$f(\al \bar a_n^- u +\be \bar a^+_n eu)=0.$$
Thus, \eqref{r} follows from $fu=0$ and $hu=ru$.

In the following Proposition we show that \eqref{r} can not be satisfied for all $n$.

\begin{Prop}
  Let $u\in \bar\repX'_{s,p}$ be a nonzero vector satisfying $hu=ru$
  with $r>0$.  Then there exists $n\in\oZ$ such that $r\bar a_n^- u
  +\bar a^+_n eu\ne 0.$
\end{Prop}

\begin{proof}
  We use the vertex operator realization of $\bar\repX'_{s,p}$.
  Namely, we consider the space $\h$ with a
  fixed nondegenerate form $(\cdot,\cdot)$ and an orthogonal basis
  $e_1,\dots, e_{p+2}$ such that
  $$(e_1,e_1)= \dots =(e_{p+1},e_{p+1})=1,\ (e_{p+2},e_{p+2})=-1.$$
  Let $v_+, v_-, v_1,\dots,v_{p-1}\in\mathbb{R}^{p+2}$ be a set of
  linearly independent vectors with
  $$(v_+,v_+)=(v_+,v_-)=(v_-,v_-)=\frac{p}{2},\ (v_\pm, v_i)=i,\
  (v_i,v_j)=2\min(i,j).$$
  For example, one can fix
\begin{gather*}
  v_i=\sqrt{2}(e_1+\dots +e_i),\ i=1,\dots, p-1,\\
  v_+=\frac{1}{\sqrt{2}}(e_1+\dots+e_{p-1})+e_p +\frac{1}{\sqrt{2}} e_{p+2}, \\
  v_-=\frac{1}{\sqrt{2}}(e_1+\dots+e_{p-1})+e_{p+1}-\frac{1}{\sqrt{2}}
  e_{p+2}.
\end{gather*}

Let
$$\widehat\h=\h\otimes\mathbb{C}[t,t^{-1}]\oplus \mathbb{C} K$$
be the
multi-dimensional Heisenberg algebra with the bracket
$$[\al\T t^i, \be\T t^j]=i\delta_{i,-j}(\al,\be) K,\ [K,\al\T t^i]=0,\
\al,\be\in\h.$$
For $\al\in\h$, let $\pi_\al$ be the Fock module with
highest--weight $\la$.  This module is generated from the highest
weight vector $|\al\rangle$ such that
$$(\be\T t^n) |\al\rangle=0,\ n>0; \qquad (\be\T 1)
|\al\rangle=(\be,\al) |\al\rangle; \qquad K|\al\rangle=|\al\rangle.$$
The $q$-degree on $\pi_\al$ is defined by
\begin{equation}
\label{defqdeg}
\deg_q |\al\rangle= \frac{(\al,\al)}{2},\quad \deg_q (\be\T t^n)=-n.
\end{equation}
We also recall the vertex operators $\Gamma_\al(z)$ acting from
$\pi_\be$ to $\pi_{\al+\be}$ with the Fourier decomposition
$$\Gamma_\al(z)=\sum_{n\in\mathbb{Z}} \Gamma_\al(n) z^{-n-(\al,\al)/2}.$$
We need two properties of vertex operators:
\begin{gather}
\label{one}
[\al\T t^n,\Gamma_\be(z)]=(\al,\be)z^n \Gamma_\be(z),\\
\label{two}
\Gamma_\al(z) \Gamma_\be(w) \sim (z-w)^{(\al,\be)}.
\end{gather}

We also need the following statement.  There exists an element
$\al_s\in\h$ such that
\begin{gather}
    \label{vpm}
    \Gamma_{v_\pm}(j) |\al_s\rangle=0,\quad j>-\frac{3p-2s}{4},\\
    \label{vn}
    \Gamma_{v_n}(j) |\al_s\rangle=0,\quad j>\left\{
    \begin{array}{l}
     -n,\quad n<s,\\
     -2n+s-1,\quad n\geq s.
    \end{array}
\right.
\end{gather}

We let $Vert_s$ denote the space generated from the vector
$|\al_s\rangle$ with all modes of the vertex operators
$\Gamma_{v_\pm}(z)$, $\Gamma_{v_n}(z)$.  Comparing the definition of
$\bar\repX'_{s,p}$ and formulas $(\ref{vpm})$, $(\ref{vn})$,
$(\ref{two})$, we obtain that the proof of the Proposition follows from
the Lemma below.
\end{proof}

\begin{Lemma}
Let $u_1$, $u_2$ be two vectors from some Fock modules. Suppose
\begin{equation}
\label{r1}
r\Gamma_{v_+}(n)u_1+ \Gamma_{v_-}(n)u_2=0
\end{equation}
for all $n$. Then $u_1=u_2=0$.
\end{Lemma}
\begin{proof}
To prove our lemma we apply an operator $\al\otimes t^i$ to both sides of
\eqref{r}. Note that for $i$ big enough
$$(\al\otimes t^i)u_1= (\al\otimes t^i)u_2.$$
Now using $(\ref{one})$
we obtain from $(\ref{r1})$ that for all $n\in\oZ$ and all $\al\in\h$
$$
(\al,v_-) r\Gamma_{v_-}(n+i)u_1+ (\al, v_+)\Gamma_{v_+}(n+i)u_2=0.
$$
Because of the nondegeneracy of $(\cdot,\cdot)$ we conclude that
for all $n$
$$\Gamma_{v_-}(n)u_1=\Gamma_{v_+}(n)u_2=0.$$
This gives $u_1=u_2=0$. Lemma is proved.
\end{proof}

\begin{Prop}
\label{sl2}
$\ch \repX_{s,p}[r]= \ch \bar\repX_{s,p}[r]=\ch \bar\repX'_{s,p}[r]$.
\end{Prop}

\begin{cor}
$\ch \repX_{s,p}=\ch \bar\repX'_{s,p}[r]$ and therefore the statement
of the Theorem $\ref{thm:main}$ is satisfied.
\end{cor}
\begin{proof}
Follows from Lemma $\ref{chx'}$ and Proposition $\ref{sl2}$.
\end{proof}

\section{Conclusion}
From the results of the paper we can obtain the following description
of $\algA(p)$ irreducible representations~$\repX_{s,p}$. We know that
$\repX_{s,p}$ is generated from the vacuum vector $\ket{s,p}$
satisfying the defining relations~\eqref{cond}.  This means that
$\repX_{s,p}$ is induced from trivial representation of the subalgebra
generated by $a^\pm_i$ with~$i=\frac{3p-2s}{4}+n$, $n\in\oN$.
In~\cite{[FLT]}, fermionic formulas for irreducible representations of
$1$-dimensional lattice vertex operator algebras and fermionic
formulas for coinvariants in irreducible representations with respect
to different subalgebras were obtained. These give graded (or quantum)
version of the Verlinde formula for $1$-dimensional lattice vertex
operator algebras.  In this paper generalization of some results
of~\cite{[FLT]} are obtained.

The $\algA(p)$ representation category is equivalent to the
representation category $\mathfrak{C}(p)$ of the small quantum
$s\ell(2)$ group $U_q(s\ell(2))$ with $q=e^{\frac{i\pi}{p}}$.  This
group differs from the quantum group $\overline{\mathscr{U}}_{q}
s\ell(2)$ from~\cite{[FGST2]} by the relation $K^p=1$. The
coinvariants in $\algA(p)$ irreducible representations can be
described in terms of~$\mathfrak{C}(p)$. Therefore the next natural
step of investigations can be obtaining of fermionic formulas for
coinvariants, which gives $q$-versions for multiplicities of a given
indecomposable representation in a tensor product of irreducible
representations.

The close related to the previous direction of investigations is a
monomial basis constructed in terms of $a^\pm(z)$-modes like
in~\cite{FJMMT}. These basises allows to establish a contact with some
RSOS-like models as in~\cite{Foda:1998tx}.

\medskip

For applications to percolation type models, a generalization of the
results of this paper to~$(p,p')$ models~\cite{[FGST3]} and especially
to $(2,3)$ model would be very useful.

\subsubsection*{Acknowledgments}
%We are grateful to\dotfill

This paper was supported by the RFBR Grant 07-01-00523.  The work of
EF and IYuT was supported in part by LSS-4401.2006.2 grant.  The work
of EF was supported in part by RFBR grants 06-01-00037 and
07-02-00799.  The work of BF was supported in part by NSch 6358.2006.2
grant and the RFBR grants 05-01-00007 and 05-01-02934.  The work of
IYuT was supported in part by the RFBR Grants 05-02-17217, $05-01-02934-YaF_a$ 
and the ``Dynasty'' foundation.

\end{document}